\documentstyle[12pt]{article}
\def\ie{{\em i.e.}}
\def\eg{{\em e.g.}}

\def\beq{\begin{equation}}
\def\eeq{\end{equation}}

\catcode`\@=11 
\def\coeff#1#2{{\textstyle{#1\over #2}}}

\def\lsim{\mathrel{\mathpalette\@versim<}}
\def\gsim{\mathrel{\mathpalette\@versim>}}
\def\@versim#1#2{\vcenter{\offinterlineskip
    \ialign{$\m@th#1\hfil##\hfil$\crcr#2\crcr\sim\crcr } }}
\def\etal{{\em et. al.}}
\def\JL{J. L. Lopez}
\def\DVN{D. V. Nanopoulos}

\def\t1{{\tilde 1}}

\def\GeV{\,{\rm GeV}}
\def\TeV{\,{\rm TeV}}

\def\to{\rightarrow}

\def\NPB#1#2#3{Nucl. Phys. B {\bf#1} (19#2) #3}
\def\PLB#1#2#3{Phys. Lett. B {\bf#1} (19#2) #3}

\def\PRD#1#2#3{Phys. Rev. D {\bf#1} (19#2) #3}
\def\PRL#1#2#3{Phys. Rev. Lett. {\bf#1} (19#2) #3}

\begin{document}
\begin{flushright}
\baselineskip=12pt
{CERN-TH.7335/94}\\
{CTP-TAMU-33/94}\\
{ACT-11/94}\\
\end{flushright}

\begin{center}
{\Huge\bf New constraints on supergravity models from $b\to s\gamma$\\}
\vglue 1cm
{JORGE L. LOPEZ$^{(a),(b)}$, D. V. NANOPOULOS$^{(a),(b),(c)}$,
XU~WANG$^{(a),(b)}$, and A.~ZICHICHI$^{(d)}$\\}
\vglue 0.4cm
{\em $^{(a)}$Center for Theoretical Physics, Department of Physics, Texas A\&M
University\\}
{\em College Station, TX 77843--4242, USA\\}
{\em $^{(b)}$Astroparticle Physics Group, Houston Advanced Research Center
(HARC)\\}
{\em The Mitchell Campus, The Woodlands, TX 77381, USA\\}
{\em $^{(c)}$CERN Theory Division, 1211 Geneva 23, Switzerland\\}
{\em $^{(d)}$CERN, 1211 Geneva 23, Switzerland\\}
\baselineskip=12pt

\vglue 1cm
{ ABSTRACT}
\end{center}
\vglue 0.5cm
{\rightskip=3pc
 \leftskip=3pc
\noindent
We perform a detailed study of the constraints from $b\to s\gamma$ on a large
class of supergravity models, including generic four-parameter supergravity
models, the minimal $SU(5)$ supergravity model, and $SU(5)\times U(1)$
supergravity. For each point in the parameter spaces of these models
we obtain a range of $B(b\to s\gamma)$ values which should conservatively
account for the unknown next-to-leading-order QCD corrections. We then
classify these points into three categories: ``excluded" points have ranges
of $B(b\to s\gamma)$ which do not overlap with the experimentally allowed
range, ``preferred" points have $B(b\to s\gamma)$ ranges which overlap with
the Standard Model prediction, and ``Ok" points are neither ``excluded" nor
``preferred" but may become ``excluded" should new CLEO data be consistent
with the Standard Model prediction. In {\em all} cases we observe a strong
tendency for the ``preferred" points towards one sign of the Higgs mixing
parameter $\mu$. For the opposite sign of $\mu$ there is an upper bound on
$\tan\beta$: $\tan\beta\lsim25$ in general, and $\tan\beta\lsim6$ for the
``preferred" points. We conclude that new CLEO data will provide a decisive
test of supergravity models.}

\vspace{0.5cm}
\begin{flushleft}
\baselineskip=12pt
{CERN-TH.7335/94}\\
{CTP-TAMU-33/94}\\
{ACT-11/94}\\
June 1994
\end{flushleft}

\vfill\eject
\setcounter{page}{1}
\pagestyle{plain}
\baselineskip=14pt

\section{Introduction}
Following the precise verification of the unification of the gauge couplings
in supergravity unified models \cite{EKN}, there has been a rekindling of
interest in low-energy supersymmetric models and their experimental
consequences. Supersymmetry can be tested in high-energy collider experiments
through the direct production of supersymmetric particles, and in dedicated
low-energy experiments through the precise measurement of rare Standard Model
loop-level processes. Because of the large energies required to produce real
supersymmetric particles, high-precision experiments are likely to
explore the parameter space of supersymmetric models in a much deeper although
indirect way for some time to come. In the context of $B$-physics, and
the $b\to s\gamma$ process in particular, there appear to be great prospects
for a thorough study of this mode at the upgraded CLEO/CESR facility and
at planned $B$-factories at SLAC and KEK.

At present there is an experimental upper bound $B(b\to
s\gamma)<5.4\times10^{-4}$, and the observation of the $B\to K^*\gamma$ process
imposes a conservative lower bound $B(b\to s\gamma)>0.6\times10^{-4}$
\cite{Thorndike}. More precise measurements are expected to be announced soon.
Despite the availability of precise experimental data, the large \cite{QCDold}
and only partially calculated \cite{QCDnew} QCD corrections to this process
make it unclear \cite{AG,Buras,Ciuchini} that one can use the data effectively
to test the Standard Model or constrain models of new physics. Nonetheless, two
alternatives appear possible: either the data deviate greatly from the Standard
Model prediction or the data agree well with the Standard Model prediction.
Since supersymmetric models predict a wide range of $b\to s\gamma$ rates (above
and below the Standard Model prediction)
\cite{Bertolini,Oshimo,BG,LNP,GNg,Okada,Diaz,Borzumati,BV}
in either case important restrictions on the parameter space will follow.

Generic low-energy supersymmetric models, like the Minimal Supersymmetric
Standard Model (MSSM), are plagued by a large number of parameters (at least
twenty), whose arbitrary tuning allows a wide range of possible experimental
predictions. This fact has discouraged many experimentalists, since any
experimental limit would appear to be always avoidable by suitable tuning of
the many parameters. These many-parameter models lack theoretical motivation,
whereas theoretically well-motivated supersymmetric models have invariably much
fewer parameters and can be straightforwardly falsified. In this context we
consider unified supergravity models with universal soft supersymmetry breaking
and radiative electroweak symmetry breaking, thus restricting the number of
parameters to four. To sharpen the predictions even more, we also consider
string-inspired no-scale $SU(5)\times U(1)$ supergravity, which can be
described in terms of two parameters, or even only one parameter in its strict
version.

Calculations of $B(b\to s\gamma)$ in supergravity models have been performed
in Refs.~\cite{Bertolini,LNP,Diaz,Borzumati,BV} under various assumptions,
such as radiative electroweak breaking using the tree-level Higgs potential
\cite{Bertolini,Diaz,Borzumati}, or imposing relations among
the supersymmetry breaking parameters (such as $B=A-1$
\cite{Bertolini,Diaz,Borzumati,BV}, or $B=2$ \cite{BV}, or $m_0=A=0$
\cite{LNP}). In this paper we revisit the $B(b\to s\gamma)$ calculation in
string-inspired no-scale $SU(5)\times U(1)$ supergravity and in the minimal
$SU(5)$ supergravity model. For a direct comparison with other possible
supergravity models, we extend our calculations to a large class of generic
four-parameter supergravity models.

It is found that $B(b\to s\gamma)$ can be enhanced in supersymmetric
models in the presence of light sparticles, and more importantly for large
values of the ratio of Higgs vacuum expectation values ($\tan\beta$). Indeed,
large values of $\tan\beta$ enhance the down-type quark and charged lepton
Yukawa couplings, in particular $\lambda_b\propto1/\cos\beta\sim\tan\beta$
for large $\tan\beta$. (An analogous effect also enhances the supersymmetric
contribution to the anomalous magnetic moment of the muon, through a large muon
Yukawa coupling \cite{g-2}.) We find that when the various decay amplitudes
have the same sign, this enhancement results in new upper bounds on
$\tan\beta$, assuming a supersymmetric spectrum below the TeV scale.

This paper is organized as follows. In section~\ref{models}
we describe briefly the models under study. In section~\ref{formula} we
discuss the expression used to calculate $B(b\to s\gamma)$ and the various
uncertainties involved. In section~\ref{results} we present and discuss our
results, and obtain new upper bounds on $\tan\beta$ using the present
experimentally allowed range. We also delineate the regions of parameter space
which are consistent with the Standard Model prediction, and explore the
$m_t$-dependence of our results. These show a strong preference for one sign of
the Higgs mixing parameter $\mu$. Finally in section~\ref{conclusions} we
summarize our conclusions.

\section{The supergravity models}
\label{models}
We consider unified supergravity models with universal soft supersymmetry
breaking and radiative breaking of the electroweak symmetry. At the unification
scale the models are described by four soft-supersymmetry-breaking parameters:
the universal gaugino mass ($m_{1/2}$), the universal scalar mass ($m_0$), the
universal trilinear scalar coupling ($A$), and the bilinear scalar coupling
($B$); and by four superpotential couplings: the Higgs mixing parameter $\mu$,
and the third-generation Yukawa couplings ($\lambda_t,\lambda_b,\lambda_\tau$).
At low energies there also arises the
ratio of Higgs vacuum expectation values ($\tan\beta$). The renormalization
group equations connect the values of the parameters at the low- and
high-energy scales. These nine parameters reduce to five by the imposition of
a good minimum of the one-loop effective Higgs potential at the electroweak
scale (one condition for each of the two real scalar Higgs fields; $|\mu|$ and
$B$ are determined), and by trading the set
$(\lambda_t,\lambda_b,\lambda_\tau,\tan\beta)$ for
$(m_t,m_b,m_\tau,\tan\beta)$ and using the known values of $m_b$ and $m_\tau$.
Once the five independent parameters ($m_{1/2},m_0,A,m_t,\tan\beta$) are
specified, one can obtain the whole low-energy spectrum and enforce all the
known bounds on sparticle and Higgs-boson masses. We note that unlike
Refs.~\cite{Bertolini,Diaz,Borzumati,BV}, we do not impose the additional
restriction $B=A-1$ at the unification scale. For recent reviews of this
procedure see \eg, Ref.~\cite{reviews}.

For a fixed value of the top-quark mass, we therefore have a family of
four-parameter supersymmetric models. We further specify this set as
$(m_{\chi^\pm_1},\xi_0,\xi_A,\tan\beta)$, where $m_{\chi^\pm_1}\propto m_{1/2}$
and we have defined $\xi_0\equiv m_0/m_{1/2}$ and $\xi_A=A/m_{1/2}$. This type
of ratios of soft-supersymmetry-breaking parameters occur naturally in various
string-inspired supersymmetry breaking scenarios (see below). In what follows
we consider continuous values of $m_{\chi^\pm_1}$ and a grid of values for
the other three parameters: $\tan\beta=2-40$ (in steps of 2);
$\xi_0=0,1,2,5,10$; $\xi_A=0,+\xi_0,-\xi_0$. For reference, the average squark
mass is related to the gluino mass as follows:
$m_{\tilde q}\approx (m_{\tilde g}/2.9)\sqrt{6+\xi^2_0}$, \ie,
$m_{\tilde q}\approx(0.8,0.9,1.1,1.9,3.6)m_{\tilde g}$ for $\xi_0=0,1,2,5,10$.

Within this class of four-parameter models we also consider the minimal
$SU(5)$ supergravity model, where one finds two additional
constraints: a sufficiently long proton lifetime \cite{LNP+LNPZ+AN} and a
sufficiently small neutralino relic density \cite{cosmo}. In what follows,
the proton lifetime has been calculated assuming that the Higgs-triplet
mass does not exceed ten times the unification scale (which implies
$\tan\beta\lsim10$ and $\xi_0\gsim4$). These combined constraints restrict
the parameter space to $m_{\chi^\pm_1}\lsim120\GeV$ and $m_{\tilde
g}\lsim400\GeV$.

An even more predictive class of supersymmetric models is obtained by
considering no-scale $SU(5)\times U(1)$ supergravity \cite{Faessler}. In this
case unification occurs at the string scale $M_{\rm string}\sim10^{18}\GeV$,
and one assumes string-inspired relations among the soft-supersymmetry-breaking
parameters resulting in two scenarios:
(i) the {\em moduli} scenario with $m_0=A=0$ \cite{EKNI+II+Lahanas,LN}, and
(ii) the {\em dilaton} scenario with $m_0={1\over\sqrt{3}}m_{1/2}$,
$A=-m_{1/2}$ \cite{KL}. In either case
one obtains two-parameter models ($m_{\chi^\pm_1},\tan\beta$). These two
scenarios have been studied in detail in Refs.~\cite{LNZI,LNZII} respectively.
One finds that the proton decay and cosmological constraints are satisfied
automatically in both scenarios. We also consider a ``strict" version of these
scenarios, where the parameter $B$ at the unification scale is also specified:
(i) $B(M_U)=0$ (moduli scenario or ``strict no-scale"), and (ii) $B(M_U)=2m_0$
(dilaton scenario or ``special dilaton"). The specification of $B(M_U)$
allows one to determine $\tan\beta$: $\tan\beta=\tan\beta(m_{\chi^\pm_1})$,
\ie, one obtains one-parameter models. Moreover, the sign of $\mu$ is
restricted to be negative in both these cases.\footnote{In the ``strict
no-scale" case both signs of $\mu$ are in principle allowed. However, $\mu>0$
can occur only for $m_t\lsim135\GeV$ which appears quite disfavored by the
latest experimental information on the top-quark mass.}

In our calculations we have taken the ``running" top-quark mass to be
$m_t\equiv m_t(m_t)=150\GeV$. (In section~\ref{mtdep} we comment on the
$m_t$-dependence of
our results.) This parameter is related to the experimentally observable
``pole" mass by~\cite{GBGS}
\begin{equation}
m_t^{\rm pole}=m_t(m_t)\left[1+\coeff{4}{3}{\alpha_s(m_t)\over\pi}
+K_t\left({\alpha_s(m_t)\over\pi}\right)^2\right]
\label{pole}
\end{equation}
where
\begin{equation}
K_t=16.11-1.04\sum_{m_{q_i}<m_t}\left(1-{m_{q_i}\over m_t}\right)\approx11.
\end{equation}
Thus we obtain $m^{\rm pole}_t\approx 1.07 m_t$, and our choice $m_t=150\GeV$
corresponds to $m^{\rm pole}_t=160\GeV$, which is in good agreement with the
latest fit to all data $m_t=162\pm9\GeV$ \cite{EFL}.

When large values of $\tan\beta$ are allowed, we also consider the constraints
from the anomalous magnetic moment of the muon. This constraint is not as
restrictive as that from $B(b\to s\gamma)$, but nonetheless can exclude
additional points in parameter space for $\mu<0$. Also, in the generic
four-parameter models, the cosmological constraint of a not-too-large
neutralino relic density becomes quite restrictive for $\xi_0\gsim2$ and
not-too-large values of $\tan\beta$ \cite{ssmdm}. This constraint is also
imposed below.

In what follows we will present our results in a way that makes apparent the
impact of $B(b\to s\gamma)$ on the parameter space, irrespective of any
additional applicable constraints, and also when all constraints are combined.

\section{Formula and uncertainties}
\label{formula}
We use the following leading-order (LO) expression for the branching ratio
\cite{BG}
\begin{equation}
{B(b\to s\gamma)\over B(b\to ce\bar\nu)}={6\alpha\over\pi}
{|V^*_{ts}V_{tb}|^2\over|V_{cb}|^2}{1\over g(m_c/m_b)}
\left[\eta^{16/23}A_\gamma
+\coeff{8}{3}(\eta^{14/23}-\eta^{16/23})A_g+C\right]^2,
\label{expression}
\end{equation}
where $\eta=\alpha_s(M_Z)/\alpha_s(Q)$, $g$ is the phase-space factor
$g(x)=1-8x^2+8x^6-x^8-24x^4\ln x$, $C=C(Q)$ is a QCD correction factor,
and $Q$ is the renormalization scale. The $A_\gamma,A_g$ are the coefficients
of the effective $bs\gamma$ and $bsg$ penguin operators evaluated at the scale
$M_Z$. These coefficients receive five contributions \cite{Bertolini} from: the
$t-W^\pm$ loop, the $t-H^\pm$ loop, the $\chi^\pm_{1,2}-\tilde t_{1,2}$ (the
``chargino" contribution), the $\tilde g-\tilde q$ loop (the ``gluino"
contribution), and the $\chi^0_i-\tilde q$ loop (the ``neutralino"
contribution). The gluino and neutralino contributions are much smaller than
the chargino contribution \cite{Bertolini,Borzumati} and are neglected in the
following. In fact, it is when the chargino contribution becomes large (for
light sparticles and large $\tan\beta$) that $B(b\to s\gamma)$ can greatly
deviate from the Standard Model result. As only the top-squarks ($\tilde
t_{1,2}$) are significantly split, in the chargino contribution we ignore the
other squark splittings. The top-squark splitting affects the magnitude of the
chargino contribution, and for fixed $m_t$ it depends mostly on the parameter
$A$ (or at low energies on $A_t$). Expressions for $A_\gamma,A_g$ can be found
in Ref.~\cite{BG}.

As is, this expression for $B(b\to s\gamma)$ is subject to partially unknown
next-to-leading-order (NLO) QCD corrections. It has been recently shown
\cite{Buras,Ciuchini} that the magnitude of the NLO corrections can be
estimated by allowing the renormalization scale $Q$ to vary between
$m_b/2$ and $2m_b$. A complete NLO calculation would yield and expression
with a much milder $Q$ dependence, with the expectation that the NLO value
for $B(b\to s\gamma)$ would be obtained from the LO expression for a choice
of $Q$ in the $m_b/2\to 2m_b$ interval. Therefore, in what follows we use
the LO expression~\footnote{In Eq.~(\ref{expression}) we have removed from the
denominator the QCD correction factor for the semileptonic decay that was used
in our previous analyses \cite{LNP}. This factor has to be removed in order to
obtain a true LO expression, without any NLO contributions.}
in Eq.~(\ref{expression}) and obtain a range of values for each point in
parameter space by taking $m_b/2<Q<2m_b$, with $m_b=4.65\GeV$.
Consistent with this procedure we use the one-loop approximation to the running
of the strong coupling, \ie,
$\eta=\alpha_s(M_Z)/\alpha_s(Q)=1-(22/3)(\alpha_s(M_Z)/2\pi)
\ln(M_Z/Q)$ (we take $\alpha_s(M_Z)=0.120$). In evaluating
Eq.~(\ref{expression}) we also take
$|V^*_{ts}V_{tb}|^2/|V_{cb}|^2=0.95\pm0.04$, $m_c/m_b=0.316\pm0.013$, and
$B(b\to ce\bar\nu)=10.7\%$ \cite{Ciuchini}. Finally $C=C(Q)=\sum_{i=1}^8
b_i\,\eta^{d_i}$, with the $b_i,d_i$ coefficients given in Ref.~\cite{BG}. For
$Q=m_b/2,m_b,2m_b$ we obtain $\eta=0.486,0.583,0.680$ and
$C=-0.208,-0.160,-0.117$. Following the above procedure, but keeping only the
$t-W^\pm$ contribution to $A_\gamma,A_g$ we obtain the following range for the
Standard Model contribution
\begin{equation}
B(b\to s\gamma)_{\rm SM}=(1.97-3.10)\times10^{-4}\, .
\label{SM}
\end{equation}

The above discussion of QCD corrections implicitly assumes that only two mass
scales are involved in the problem: the high electroweak scale and the
low $b$-quark mass scale. In practice the supersymmetric particles have
a spectrum that can be spread above or below the electroweak scale. Recently
there has appeared the first study of QCD corrections to $B(b\to s\gamma)$ in
the supersymmetric case \cite{Anlauf}, which shows that the running from the
electroweak scale down to the $b$-quark mass scale gives the largest QCD
correction. However, a proper treatment of the supersymmetric spectrum at the
high scale may produce non-negligible effects. We hope that the scale
uncertainty that we have introduced above is large enough to effectively
encompass the supersymmetric high scale uncertainty as well.

\section{Results and discussion}
\label{results}
We now present the results of the calculations of $B(b\to s\gamma)$ for
the various supergravity models under consideration. We start by revisiting
the results in $SU(5)\times U(1)$ supergravity and the minimal $SU(5)$
supergravity model, and then extend our calculations to the generic
four-parameter models described in section~\ref{models}.
\subsection{SU(5)xU(1) supergravity}
\label{flipped}
The calculated values of $B(b\to s\gamma)$ in the moduli and dilaton scenarios
are shown in Fig.~\ref{bsgnsc} and Fig.~\ref{bsgkl} respectively, for selected
values of $\tan\beta$; consistency of the model entails an upper limit of
$\tan\beta\lsim26\,(40)$ in the moduli (dilaton) scenario. In these and other
figures showing values of $B(b\to s\gamma)$ we take $Q=m_b$ as a ``central"
value. The full uncertainty range is considered when discussing whether given
points in parameter space are excluded or not. In
Figs.~\ref{bsgnsc},\ref{bsgkl} the arrows point into the experimentally allowed
region and the dashed lines (``SM") delimit the Standard Model prediction. For
$\mu>0$ the values of $B(b\to s\gamma)$ increase steadily with increasing
$\tan\beta$ and eventually fall outside the experimentally allowed region for
all values of $m_{\chi^\pm_1}$. (The largest values of $m_{\chi^\pm_1}$ shown
correspond to $m_{\tilde q},m_{\tilde g}\gsim1\TeV$.)

For $\mu<0$ the $\tan\beta$-dependence is different. One sees that $B(b\to
s\gamma)$ can be suppressed much below the
Standard Model result. This behavior was first noticed in Ref.~\cite{LNP}
and simply shows that the various amplitudes and QCD correction factors
in Eq.~(\ref{expression}) conspire to produce a cancellation. This phenomenon
has been since explained in Refs.~\cite{GNg,BV}. The idea is that the chargino
contribution to $A_\gamma$ can have the same sign (negative) or opposite sign
(positive) compared to the $t-W^\pm$ and $t-H^\pm$ contributions which are
always negative. In fact, the sign of the chargino contribution is determined
by the product $\theta_{\tilde t}\mu$: positive for $\theta_{\tilde t}\mu<0$
and negative for $\theta_{\tilde t}\mu>0$
\cite{GNg}.\footnote{Our sign convention for $\mu$ is opposite to that in
Ref.~\cite{GNg}, but the same as that in Ref.~\cite{BV}.}
Here $\theta_{\tilde t}$ is the top-squark mixing angle, $\theta_{\tilde
t}\approx\pi/4$ in this approximation to the chargino contribution \cite{BG}.
Therefore, constructive interference occurs for $\mu>0$ and destructive
interference occurs for $\mu<0$, as evident in Figs.~\ref{bsgnsc},\ref{bsgkl}.
Fig.~\ref{bsgkl} also shows that despite the destructive interference, for
$\mu<0$ and sufficiently large values of $\tan\beta$, an enhancement can occur
for light chargino masses because the chargino contribution overwhelms the
other two contributions.

To better appreciate the impact of the present experimental limit on $B(b\to
s\gamma)$, we classify points in parameter space into three categories:
\begin{itemize}
\item {\em Excluded points} have $B(b\to s\gamma)$ ranges which do not overlap
with the present experimental allowed range (\ie, $(0.6-5.4)\times10^{-4}$).
This conservative constraint (\ie, use of theoretical ranges rather than
central values) is imposed so that our excluded points are not dependent on
presently unknown NLO QCD corrections.
\item {\em Preferred points} have $B(b\to s\gamma)$ ranges which
overlap with the Standard Model range (\ie, $(1.97-3.10)\times10^{-4}$).
\item {\em Ok points} are neither ``excluded" nor ``preferred", and may become
``excluded" if new CLEO data are consistent with the Standard Model prediction.
\end{itemize}
The two-dimensional parameter spaces for the moduli and dilaton scenarios are
shown in Figs.~\ref{PSnsc} and \ref{PSkl} respectively. In these figures
crosses (`$\times$') represent ``excluded" points, diamonds (`$\diamond$')
represent ``preferred" points, and dots (`$\cdot$') represent ``Ok" points.
The vertical dashed line at $m_{\chi^\pm_1}=100\GeV$ in these and following
parameter space plots indicates the approximate reach of LEPII for chargino
masses. As anticipated, there are many ``excluded" points for $\mu>0$,
especially as $\tan\beta$ gets large. Also, the ``preferred" points occur
largely for $\mu<0$, signalling the ability of the $B(b\to s\gamma)$ constraint
to possibly select the sign of $\mu$.

For sufficiently large values of $\tan\beta$, the anomalous magnetic moment
of the muon $(g-2)_\mu$ can be constraining as well. We have calculated
this quantity as in Ref.~\cite{g-2} and denote excluded points (\ie, points
for which ${1\over2}(g-2)^{\rm susy}_\mu$ falls outside the experimentally
allowed range of $-13.2\times10^{-9}\to 20.8\times10^{-9}$ \cite{g-2}) by
plusses (`$+$') in Figs.~\ref{PSnsc},\ref{PSkl}. We have only shown these
excluded points for $\mu<0$, since for $\mu>0$ the $(g-2)_\mu$ constraint does
not exclude any points which are not excluded by the $B(b\to s\gamma)$
constraint. The effect of this constraint is most noticeable in the dilaton
scenario (Fig.~\ref{PSkl}) where ``preferred" points which are thusly excluded
appear as the overlap of a plus sign and a clear diamond (\ie, ``filled"
diamonds) for $\tan\beta=10-32$ and $m_{\chi^\pm_1}\lsim100\GeV$. Points
excluded by both $B(b\to s\gamma)$ and $(g-2)_\mu$ appear as the overlap of
a plus sign and a cross, \ie, as asterisks (`$\ast$').

Overall, the $B(b\to s\gamma)$ constraint is quite restrictive. Statistically
we have the following distribution of fractions of parameter space:
\begin{center}
\begin{tabular}{|c||c|c||c|c|}\hline
&\multicolumn{2}{c||}{Moduli}&\multicolumn{2}{c|}{Dilaton}\\ \hline
&$\mu>0$&$\mu<0$&$\mu>0$&$\mu<0$\\ \hline
Excluded&48\%&7\%&73\%&21\%\\
Preferred&9\%&66\%&6\%&33\%\\
Ok&43\%&27\%&21\%&46\%\\ \hline
\end{tabular}
\end{center}
Note that should the new CLEO data be consistent with the Standard Model
prediction, then only the ``preferred" points would survive. For $\mu>0$
this is  $9\,(6)\%$ of all the points in parameter space in the moduli
(dilaton) scenario, whereas for $\mu<0$ one would still have
${2\over3}\,({1\over3})$ of the points allowed -- an overwhelming inclination
towards $\mu<0$.

{}From Figs.~\ref{PSnsc},\ref{PSkl} one can also see that for $\mu<0$ there
is an upper bound on $\tan\beta$: $\tan\beta\lsim25$. For the subset of
``preferred" points this bound drops to $\tan\beta\lsim6$. We should add that
these bounds can be evaded by demanding sufficiently heavy sparticles (in the
multi-TeV range), but then supersymmetry would lose most of its motivation.
There are no analogous bounds for $\mu<0$.

Note that the ``preferred" points for $\mu>0$ entail a supersymmetric
spectrum inaccessible to direct searches at LEPII (\ie,
$m_{\chi^\pm_1}\gsim150\GeV$, $m_{\tilde g}\approx m_{\tilde q}\gsim500\GeV$).
In contrast, some of the ``preferred" points for $\mu<0$ would be directly
accessible at LEPII.

In the strict no-scale case, the one-parameter models allow one to plot
$B(b\to s\gamma)$ as a function of $m_{\chi^\pm_1}$ only. These values are
shown in Fig.~\ref{bsgB}. Since in both cases $\mu<0$, the plots are
reminiscent of the $\mu<0$ plots in the general moduli and dilaton scenarios.
In Fig.~\ref{PSB} we show the corresponding one-dimensional parameter spaces,
where one can see that in the dilaton scenario there are no excluded points.
The constraint from $(g-2)_\mu$ is ineffective in the dilaton scenario and
does not exclude any further points in the moduli scenario. Figure~\ref{PSB}
shows the interest of working with a theory where all predictions are given
in terms of only one parameter. For example, a measurement of $m_{\chi^\pm_1}$
would immediately determine $\tan\beta$. Also, if it is found that
$2\lsim\tan\beta\lsim12$ or $\tan\beta\gsim19$ are preferred, then this
one-parameter theory should be abandoned.

\subsection{Minimal SU(5) supergravity}
As described in section~\ref{models}, in the case of the minimal $SU(5)$
supergravity model we constrain the four-dimensional parameter space by
the requirements of proton decay and cosmology. These entail $\tan\beta\lsim10$
and $\xi_0\gsim4$. In Fig.~\ref{bsgmin} we show the calculated central values
of $B(b\to s\gamma)$ in this case.  Because the values of $\tan\beta$ are not
allowed to be large we do not expect large deviations from the Standard Model
prediction, as seen in the figure. Nonetheless, there are deviations,
especially for $\mu>0$. Given the uncertainties in the calculation of $B(b\to
s\gamma)$ described in section~\ref{formula} we find no ``excluded" points,
and the following distribution of fractions of parameter space:
\begin{center}
\begin{tabular}{|c|c|c|}\hline
&\multicolumn{2}{c|}{Minimal SU(5)}\\ \hline
&$\mu>0$&$\mu<0$\\ \hline
Excluded&0\%&0\%\\
Preferred&11\%&93\%\\
Ok&89\%&7\%\\ \hline
\end{tabular}
\end{center}
Again we see a strong tendency among the ``preferred" points towards $\mu<0$.
Moreover, for $\mu>0\,(\mu<0)$ ``preferred" points exist only for
$\tan\beta\lsim4\,(8)$.

\subsection{Generic supergravity models}
For completeness we now turn to the generic four-parameter supergravity models.
These models could be viewed as essentially $SU(5)$ supergravity models where
the constraint from proton decay is simply ignored (the cosmological constraint
is not neglected). In the spirit of Ref.~\cite{ssmdm}, we consider these models
to see if $b\to s\gamma$ could shed some light on the structure of
the soft supersymmetry breaking sector in supergravity or superstring models.
As mentioned in section~\ref{models}, we have considered continuous values of
$m_{\chi^\pm_1}$ and a grid of values for the other three parameters:
$\tan\beta=2-40$ (in steps of 2); $\xi_0=0,1,2,5,10$; $\xi_A=0,+\xi_0,-\xi_0$.
We will concentrate on $\xi_0=1,2$ since the $SU(5)\times U(1)$ scenarios
correspond approximately to $\xi_0=0,{1\over\sqrt{3}}$, and larger values of
$\xi_0$ lead to increasingly heavier sparticle masses. We will comment on
the $\xi_0=5,10$ cases later. In Figs.~\ref{bsgSSMb},\ref{bsgSSMc} we show
$B(b\to s\gamma)$ for $\xi_0=1$ and $2$ respectively, for representative values
of $\tan\beta$. These curves are for $\xi_A=0$. The effect of different choices
for $\xi_A$ is shown by the dotted lines in these figures (for the same
$\tan\beta=20$ and $\xi_0$ choice): $\xi_A<0$ enhances $B(b\to s\gamma)$ since
negative $A$ values drive $A_t$ at low energies to even more negative values
and thus larger $\tilde t_{1,2}$ splittings. Increasing $\xi_0$ from 1 to 2 has
the expected effect of decreasing $B(b\to s\gamma)$.

For fixed values of $\xi_0$ and $\xi_A$ we can plot the parameter space in
the $(m_{\chi^\pm_1},\tan\beta)$ plane, as was done in the $SU(5)\times U(1)$
supergravity case. The points in parameter space are again classified into
``excluded", ``preferred", and ``Ok" as discussed in section~\ref{flipped}.
These plots for $\xi_0=1,2$ and $\xi_A=0,+\xi_0,-\xi_0$ are shown in
Figs.~\ref{PSSSMb},\ref{PSSSMc}.

For $\xi_0=1$ the $(g-2)_\mu$ constraint excludes additional points in
parameter space for $\mu<0$. These appear as filled diamonds and plusses
(`$+$') in Fig.~\ref{PSSSMb}. The effect of $\xi_A$ discussed above is also
evident in this figure, \ie, for $\xi_A=-1$ one sees that $B(b\to s\gamma)$ is
enhanced. The island of ``preferred" points for $\mu<0$,
$m_{\chi^\pm_1}\lsim100\GeV$ and not-so-small values of $\tan\beta$ corresponds
to the curves in Fig.~\ref{bsgSSMb} which ``bounce back" for light chargino
masses. Note that most of these points are in fact excluded by the $(g-2)_\mu$
constraint. The cosmological constraint is unrestrictive for $\xi_0=1$.

For $\xi_0=2$ the $(g-2)_\mu$ constraint does not exclude any points in
parameter space which are not also excluded by the $B(b\to s\gamma)$
constraint. However, the {\em cosmological} constraint becomes important and
excludes points denoted by filled diamonds and plusses in Fig.~\ref{PSSSMc}.
Note that this constraint excludes most (if not all) of the (few) ``preferred"
points for $\mu>0$. The right boundary of the parameter space corresponds to
the squarks at $1\TeV$. The position of this boundary as a function of
$m_{\chi^\pm_1}$ depends on $\xi_A$ because $A$ affects the calculation of
$\mu$, which in turn affects the value of $m_{\chi^\pm_1}$.

In analogy with the discussion for the previous models, statistically
we have the following distribution of fractions of parameter space (excluding
the cosmological constraint):\\
\vbox{
\small
\begin{center}
\begin{tabular}{|c||ccc|ccc||ccc|ccc|}\hline
&\multicolumn{6}{c||}{$\xi_0=1$}&\multicolumn{6}{c|}{$\xi_0=2$}\\ \hline
&\multicolumn{3}{c|}{$\mu>0$}&\multicolumn{3}{c||}{$\mu<0$}
&\multicolumn{3}{c|}{$\mu>0$}&\multicolumn{3}{c|}{$\mu<0$}\\ \hline
$\xi_A$&$+1$&$0$&$-1$&$+1$&$0$&$-1$&$+2$&$0$&$-2$&$+2$&$0$&$-2$\\ \hline
Excluded&70\%&74\%&76\%&12\%&15\%&20\%&62\%&72\%&78\%&9\%&17\%&27\%\\
Preferred&4\%&4\%&4\%&53\%&43\%&36\%&7\%&6\%&5\%&51\%&36\%&29\%\\
Ok&26\%&22\%&20\%&35\%&43\%&44\%&31\%&23\%&17\%&40\%&47\%&44\%\\ \hline
\end{tabular}
\end{center}
}
In this case we again see the marked tendency of the ``preferred" points for
$\mu<0$, \ie, only few percent of the points for $\mu>0$ are ``preferred".
Including the cosmological constraint makes this tendency even more pronounced.
{}From Figs.~\ref{PSSSMb},\ref{PSSSMc} we also notice the upper bound on
$\tan\beta$ for $\mu>0$: $\tan\beta\lsim25$ in general and $\tan\beta\lsim6$
for the ``preferred" points.

For $\xi_0=5,10$ the parameter space is generally quite limited in the
$\tan\beta$ direction, \ie, $\tan\beta\lsim4\,(2)$ for $\xi_0=5\,(10)$.
Higher values of $\tan\beta$ are inconsistent with radiative electroweak
symmetry breaking. Using the tree-level scalar potential such values of
$\tan\beta$ give $\mu^2<0$. In our calculations (using the one-loop effective
potential) no minimum can be found. This limitation on $\tan\beta$ can be
circumvented for sufficiently negative values of $\xi_A$, \eg,
$\xi_A=-5\,(-10)$ for $\xi_0=5\,(10)$. In any event, the qualitative results
obtained above for the strong tendency of the ``preferred" points towards
$\mu<0$, holds also for large values of $\xi_0$. The cosmological constraint
exacerbates this tendency. For $\xi_0=10$ basically all points in parameter
space for $\mu>0$ are excluded; for $\mu<0$ a few survive.

\subsection{$m_t$ dependence}
\label{mtdep}
Our calculations above have been performed for a fixed value of $m_t$. Given
the present uncertainty on the actual value of $m_t$, it is appropriate to
explore the $m_t$ dependence of our results. In the Standard Model we obtain
the following LO ranges for $B(b\to s\gamma)_{\rm SM}$
\begin{eqnarray}
m_t&=&130\GeV,\qquad B(b\to s\gamma)_{\rm SM}=(1.77-2.89)\times10^{-4}\
;\nonumber\\
m_t&=&150\GeV,\qquad B(b\to s\gamma)_{\rm SM}=(1.97-3.10)\times10^{-4}\
;\nonumber\\
m_t&=&170\GeV,\qquad B(b\to s\gamma)_{\rm SM}=(2.15-3.28)\times10^{-4}\
;\nonumber\\
m_t&=&190\GeV,\qquad B(b\to s\gamma)_{\rm SM}=(2.31-3.44)\times10^{-4}\
.\nonumber
\end{eqnarray}
These results indicate that a measurement of $B(b\to s\gamma)$ will not
give new information on $m_t$ (since all intervals overlap), at least at the LO
level. A NLO calculation would be most useful in this regard.

We have also re-done the calculation of $B(b\to s\gamma)$ in $SU(5)\times
U(1)$ supergravity for various values of $m_t$. As expected, the $m_t$
dependence is weak. To better quantify this statement, as a function of $m_t$
we have determined the fractions of parameter space which are ``preferred":
\begin{center}
\begin{tabular}{|c||c|c||c|c|}\hline
&\multicolumn{2}{c||}{Moduli}&\multicolumn{2}{c|}{Dilaton}\\ \hline
$m_t$&$\mu>0$&$\mu<0$&$\mu>0$&$\mu<0$\\ \hline
150&9\%&66\%&6\%&33\%\\
160&8\%&66\%&5\%&32\%\\
170&8\%&66\%&4\%&31\%\\
180&2\%&67\%&1\%&28\%\\
\hline
\end{tabular}
\end{center}
Except for $m_t=180\GeV$, the fractions of parameter space which are
``preferred" are largely $m_t$ independent. The increased sensitivity
for large values of $m_t$ is interesting, although not very useful since
for $m_t\gsim190\GeV$ the allowed parameter space disappears because of the
well known phenomenon of the top-quark Yukawa coupling encountering a Landau
pole below the unification scale.

\section{Conclusions}
\label{conclusions}
We have explored a variety of supergravity models with universal soft
supersymmetry breaking at the unification scale and radiative electroweak
symmetry breaking. Accounting for the inherent uncertainties in the $B(b\to
s\gamma)$ evaluation, we have nonetheless been able to exclude a good fraction
of points in parameter space which differ significantly from the experimentally
allowed range. We have also identified ``preferred" regions of parameter
space which would be singled out should the new CLEO data be consistent with
the Standard Model prediction. These ``preferred" regions occur mostly for
one sign of $\mu(<0)$ (when the chargino contribution has opposite sign
relative to the Standard Model contribution). For the other sign of $\mu(>0)$
one can obtain upper bounds on $\tan\beta$ assuming a sparticle spectrum below
the TeV scale.

It is worth noticing that constraints from $b\to s\gamma$ disfavor
the large-$\tan\beta$ solution \cite{yuks} to the unification of the Yukawa
couplings ($\lambda_b=\lambda_\tau$) in $SU(5)$-like theories (although those
are already disfavored by proton decay constraints). In the case of
$SO(10)$-like Yukawa unification ($\lambda_t=\lambda_b=\lambda_\tau$), one
requires $\tan\beta={\cal O}(50)$ \cite{SO10}, which are also disfavored by
$b\to s\gamma$ constraints. However, in this case radiative electroweak
symmetry breaking is also difficult and it has been suggested \cite{newSO10}
that non-universal soft supersymmetry breaking scenarios may be able to solve
both these difficulties.

At the moment the $b\to s\gamma$ constraint is the most important constraint
on the parameter spaces of supergravity models. This standing is expected to
be much re-enforced when the new CLEO data are announced. At this time it will
become quite appropriate to re-examine the phenomenological implications of
the still-allowed points in parameter space for direct and indirect searches
of supersymmetric particles. We close by remarking that contrary to direct
production experiments, the $b\to s\gamma$ process can already explore regions
of parameter space with ``virtual" mass scales all the way up to the TeV scale.

\section*{Acknowledgments}
This work has been supported in part by DOE grant DE-FG05-91-ER-40633. The
work of X. W. has been supported by the World Laboratory.

\newpage

\begin{figure}[p]
\vspace{6in}
\includegraphics{bsgnsc.ps}
\caption{The calculated ``central" values of $B(b\to s\gamma)$ in no-scale
$SU(5)\times U(1)$ supergravity -- moduli scenario, for representative values
of $\tan\beta$. The arrows point into the experimentally allowed region. The
dashed lines delimit the Standard Model range.}
\label{bsgnsc}
\end{figure}
\clearpage

\begin{figure}[p]
\vspace{6in}
\includegraphics{bsgkl.ps}
\caption{The calculated ``central" values of $B(b\to s\gamma)$ in no-scale
$SU(5)\times U(1)$ supergravity -- dilaton scenario, for representative values
of $\tan\beta$. The arrows point into the experimentally allowed region. The
dashed lines delimit the Standard Model range.}
\label{bsgkl}
\end{figure}
\clearpage

\begin{figure}[p]
\vspace{5in}
\includegraphics{PSnsc.ps}
\vspace{-1.5in}
\caption{The two-dimensional parameter space of no-scale $SU(5)\times U(1)$
supergravity -- moduli scenario. Points excluded by $B(b\to s\gamma)$
are denoted by crosses (`$\times$'), those consistent with the Standard Model
prediction are denoted by diamonds (`$\diamond$'), and the rest are denoted by
dots (`$\cdot$'). Plusses (`$+$') indicate points excluded by $(g-2)_\mu$,
whereas asterisks (`$\ast$') indicate points excluded by both constraints.}
\label{PSnsc}
\end{figure}
\clearpage

\begin{figure}[p]
\vspace{5in}
\includegraphics{PSkl.ps}
\vspace{-1.5in}
\caption{The two-dimensional parameter space of no-scale $SU(5)\times U(1)$
supergravity -- dilaton scenario. Points excluded by $B(b\to s\gamma)$
are denoted by crosses (`$\times$'), those consistent with the Standard Model
prediction are denoted by diamonds (`$\diamond$'), and the rest are denoted by
dots (`$\cdot$'). Filled diamonds and plusses (`$+$') indicate points excluded
by $(g-2)_\mu$, whereas asterisks (`$\ast$') indicate points excluded by both
constraints.}
\label{PSkl}
\end{figure}
\clearpage

\begin{figure}[p]
\vspace{6in}
\includegraphics{bsgB.ps}
\caption{The calculated ``central" values of $B(b\to s\gamma)$ in strict
no-scale $SU(5)\times U(1)$ supergravity -- moduli and dilaton scenarios. The
arrows point into the experimentally allowed region. The dashed lines delimit
the Standard Model range.}
\label{bsgB}
\end{figure}
\clearpage

\begin{figure}[p]
\vspace{6in}
\includegraphics{PSB.ps}
\vspace{-0.5in}
\caption{The one-dimensional parameter space of strict no-scale $SU(5)\times
U(1)$ supergravity -- moduli and dilaton scenarios. Points excluded by $B(b\to
s\gamma)$ are denoted by crosses (`$\times$'), those consistent with the
Standard Model prediction are denoted by diamonds (`$\diamond$'), and the rest
are denoted by dots (`$\cdot$').}
\label{PSB}
\end{figure}
\clearpage

\begin{figure}[p]
\vspace{6in}
\includegraphics{bsgmin.ps}
\caption{The calculated ``central" values of $B(b\to s\gamma)$ in the
minimal $SU(5)$ supergravity model. The constraints from proton decay
and cosmology restrict the parameter space as indicated. The arrows point into
the experimentally allowed region. The dashed lines delimit the Standard Model
range.}
\label{bsgmin}
\end{figure}
\clearpage

\begin{figure}[p]
\vspace{6in}
\includegraphics{bsgSSMb.ps}
\caption{The calculated ``central" values of $B(b\to s\gamma)$ in the
generic supergravity model for $\xi_0=1$ and $\xi_A=0$, for representative
values of $\tan\beta$. The dotted curve above (below) the $\tan\beta=20$
curve for $\mu>0$ corresponds to $\xi_A=-1\,(+1)$. The arrows point into the
experimentally allowed region. The dashed lines delimit the Standard Model
range.}
\label{bsgSSMb}
\end{figure}
\clearpage

\begin{figure}[p]
\vspace{6in}
\includegraphics{bsgSSMc.ps}
\caption{The calculated ``central" values of $B(b\to s\gamma)$ in the
generic supergravity model for $\xi_0=2$ and $\xi_A=0$, for representative
values of $\tan\beta$. The dotted curve above (below) the $\tan\beta=20$
curve for $\mu>0$ corresponds to $\xi_A=-2\,(+2)$. The arrows point into the
experimentally allowed region. The dashed lines delimit the Standard Model
range.}
\label{bsgSSMc}
\end{figure}
\clearpage

\begin{figure}[p]
\vspace{5in}
\includegraphics{PSSSMba.ps}
\vspace{-1.5in}
\caption{The parameter space of the generic supergravity model for $\xi_0=1$
and (a) $\xi_A=0$, (b) $\xi_A=+1$, and (c) $\xi_A=-1$. Points excluded by
$B(b\to s\gamma)$ are denoted by crosses (`$\times$'), those consistent with
the Standard Model prediction are denoted by diamonds (`$\diamond$'), and the
rest are denoted by dots (`$\cdot$'). Filled diamonds and plusses (`$+$')
indicate points excluded by $(g-2)_\mu$, whereas asterisks (`$\ast$') indicate
points excluded by both constraints.}
\label{PSSSMb}
\vspace{-15cm}
{\large\bf(a)}
\vspace{15cm}
\end{figure}
\clearpage

\begin{figure}[p]
\vspace{2in}
\includegraphics{PSSSMbb.ps}
\vspace{-14cm}
\begin{flushleft}
{\large\bf(b)}
\end{flushleft}
\vspace{12in}
\includegraphics{PSSSMbc.ps}
\vspace{-20cm}
\begin{flushleft}
{\large\bf(c)}
\end{flushleft}
\end{figure}
\clearpage

\begin{figure}[p]
\vspace{5in}
\includegraphics{PSSSMca.ps}
\vspace{-1.5in}
\caption{The parameter space of the generic supergravity model for $\xi_0=2$
and (a) $\xi_A=0$, (b) $\xi_A=+2$, and (c) $\xi_A=-2$. Points excluded by
$B(b\to s\gamma)$ are denoted by crosses (`$\times$'), those consistent with
the Standard Model prediction are denoted by diamonds (`$\diamond$'), and the
rest are denoted by dots (`$\cdot$'). Filled diamonds and plusses (`$+$')
indicate points excluded by the {\em cosmological} constraint, whereas
asterisks (`$\ast$') indicate points excluded by both constraints. At the right
boundary of the parameter space the squark masses reach 1 TeV.}
\label{PSSSMc}
\vspace{-15cm}
{\large\bf(a)}
\vspace{15cm}
\end{figure}
\clearpage

\begin{figure}[p]
\vspace{2in}
\includegraphics{PSSSMcb.ps}
\vspace{-14cm}
\begin{flushleft}
{\large\bf(b)}
\end{flushleft}
\vspace{12in}
\includegraphics{PSSSMcc.ps}
\vspace{-20cm}
\begin{flushleft}
{\large\bf(c)}
\end{flushleft}
\end{figure}
\clearpage

\end{document}